\documentclass[12pt,onecolumn,oneside,a4paper,peerreview]{IEEEtran}
\usepackage{cite}
\usepackage{graphicx}
\usepackage{amsmath}
\usepackage{times}
\usepackage{latexsym}
\usepackage{graphicx}
\usepackage{bm}
\usepackage{amssymb}
\usepackage[center]{caption2}
\usepackage{stfloats}
\usepackage{cases}
\usepackage{subfigure}
\usepackage{array}
\usepackage{setspace}
\usepackage{fancyhdr}
\usepackage{cite}
\usepackage{color}

\newtheorem{lemma}{Lemma}

\def\proof{\noindent\hspace{2em}{\itshape Proof: }}

\newcaptionstyle{mystyle2}{%
\captionlabel $.$ \, \doublespacing \captiontext \par}
\captionstyle{mystyle2}

\begin{document}
\title{Proactive Eavesdropping in Relaying Systems}
\author{Xin Jiang,~\IEEEmembership{Student Member,~IEEE,} Hai Lin,~\IEEEmembership{Senior Member,~IEEE,} Caijun Zhong,~\IEEEmembership{Senior Member,~IEEE,} Xiaoming Chen,~\IEEEmembership{Senior Member,~IEEE,} and Zhaoyang Zhang,~\IEEEmembership{Member,~IEEE}

\thanks{Copyright (c) 2017 IEEE. Personal use of this material is permitted. However, permission to use this material for any other purposes must be obtained from the IEEE by sending a request to pubs-permissions@ieee.org.}
\thanks{Manuscript received March 23, 2017, revised April 9, 2017, accepted April 17, 2017. The associate editor coordinating the review of this paper and approving it for publication was Prof. Yong Xiang.}
\thanks{X. Jiang, C. Zhong, X. Chen and Z. Zhang are with the Institute of Information and Communication Engineering, Zhejiang University, China. (email: caijunzhong@zju.edu.cn).

H. Lin is with the Department of Electrical and Information Systems, Osaka Prefecture University, Osaka 599-8531, Japan (email: lin@eis,osakafuu.ac.jp).}}

\maketitle
\makeatletter
\newcommand{\rmnum}[1]{\romannumeral #1}
\newcommand{\Rmnum}[1]{\expandafter\@slowromancap\romannumeral #1@}
\makeatother

\begin{abstract}
This paper investigates the performance of a legitimate surveillance system, where a legitimate monitor aims to eavesdrop on a dubious decode-and-forward relaying communication link. In order to maximize the effective eavesdropping rate, two strategies are proposed, where the legitimate monitor adaptively acts as an eavesdropper, a jammer or a helper. In addition, the corresponding optimal jamming beamformer and jamming power are presented. Numerical results demonstrate that the proposed strategies attain better performance compared with intuitive benchmark schemes. Moreover, it is revealed that the position of the legitimate monitor plays an important role on the eavesdropping performance for the two strategies.
\end{abstract}


\section{Introduction}
With rapid advancements in wireless technologies, wireless communications infrastructure and services have brought great convenience to our daily lives. However, the benefits of wireless communication may be exploited by potential malicious users to commit crimes or terror attacks \cite{Y.Zou,D.Wang}. Therefore, there is a growing need for the authorized parties such as government agencies to legitimately monitor any suspicious communications to ensure public safety and prevent terrorism.

Responding to this, a new paradigm shift in wireless security by investigating how a legitimate monitor performs legitimate information surveillance, was proposed in \cite{J.Xu,J.Xu1,X.Jiang}. In particular, the authors proposed a novel approach, namely, proactive eavesdropping via jamming, where the legitimate monitor operates in a full-duplex manner, and purposely transmits jamming signals to interfere with the suspicious link while performs eavesdropping. Later in \cite{Y.Zeng}, another spoofing-relay based proactive eavesdropping approach was proposed.

Note that all the aforementioned works focus on the three-node point-to-point communication setup. Thus far, how to perform proactive eavesdropping in relaying systems remains largely an uncharted area. Motivated by this, in this paper, we propose a novel legitimate surveillance approach for dual-hop decode-and-forward (DF) relaying communication systems. Specifically, to maximize the effective eavesdropping rate as in \cite{Y.Zeng}, two strategies are designed for the legitimate monitor, which acts adaptively as an information eavesdropper, a destructive jammer or a constructive helper in the two time slots. For each strategy, the optimal jamming beamformer and jamming power are derived. Numerical results reveal that the proposed strategies achieve better performance than some intuitive benchmark schemes.

{\it Notation}: We use bold upper case letters to denote matrices, bold lower case letters to denote vectors and lower case letters to denote scalars. $||\cdot||$, $(\cdot)^\dagger$ and $\text{tr} (\cdot)$ denote Euclidean norm, conjugate transpose operator and the trace of a matrix, respectively. $\Pi_{\mathbf{X}} \triangleq \mathbf{X} \left(\mathbf{X}^\dagger \mathbf{X}\right)^{-1} \mathbf{X}^\dagger$ represents the orthogonal projection onto the column space of $\mathbf{X}$ and $\Pi_{\mathbf{X}}^{\perp} \triangleq \mathbf{I}-\Pi_{\mathbf{X}}$ denotes the orthogonal projection onto the orthogonal complement of the column space of $\mathbf{X}$.

\section{System Model}
\begin{figure}[htbp]
\centering
\includegraphics[width=2.6in,height=1.2in]{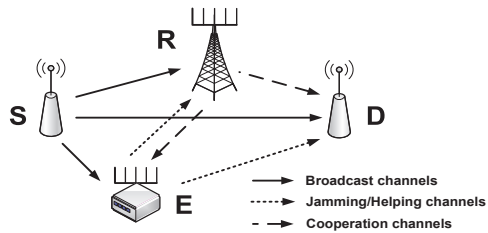}
\caption{A four-node legitimate surveillance system.}
\label{fig:fig1}
\end{figure}

As shown in Fig. \ref{fig:fig1}, we consider a four-node legitimate surveillance system, where a legitimate monitor E aims to eavesdrop the communication between a suspicious transmitter S and a suspicious receiver D, which is assisted by a DF relay R. We assume that R and E are equipped with $N$ and $M$ antennas, respectively, while S and D are equipped with a single antenna each. All nodes operate in the half-duplex mode and a direct link exists between S and D.

We adopt the time-sharing protocol \cite{A.A.Nasir}, where the entire communication consists of two time slots. The relay listens to the source transmission during the first time slot, and then forwards the decoded symbol to the destination in the second time slot. In contrast, the legitimate monitor may choose to either jam, eavesdrop, or help, depending on the channel conditions during the two time slots.

We assume that all the channel links are composed of large-scale path loss with exponent $\tau$ and statistically independent small-scale Rayleigh fading. We denote the inter-node distance of links S $\rightarrow$ R, S $\rightarrow$ D, S $\rightarrow$ E, R $\rightarrow$ E, R $\rightarrow$ D and E $\rightarrow$ D by $d_1$, $d_2$, $d_3$, $d_4$, $d_5$ and $d_6$, respectively. The corresponding small-scale fading channel coefficients are denoted by $N \times 1$ vector $\mathbf{h}_1$, scalar $h_2$, $M \times 1$ vector $\mathbf{h}_3$, $M \times N$ matrix $\mathbf{H}_4$, $1 \times N$ vector $\mathbf{h}_5$ and $1 \times M$ vector $\mathbf{h}_6$, respectively.  Quasi-static fading is assumed, such that the channel coefficients remain unchanged during each transmission block but vary independently between different blocks. Each element of these complex fading channel coefficients are circular symmetric complex Gaussian random variables with zero mean and unit variance. Channel reciprocity is also assumed.

We assume that global channel state information (CSI) is available at E\footnote{The CSI can be obtained by utilizing the methods given in paper \cite{J.Xu1,Y.Zeng}.}, while S, R and D only know the channel gains of relative suspicious channels. This assumption is practical since it is difficult for the suspicious nodes to know the existence of the legitimate monitor and hence conventional physical layer security is not applied to prevent eavesdropping.

\section{Problem Formulation}
In this section, we describe in detail the problem formulation. Depending on the particular action taken by E during the first time slot, we consider two separate scenarios.
\subsection{Strategy \uppercase\expandafter{\romannumeral1}: Jamming First}
This scenario is applicable when E is located relatively far away from S, thus E is unlikely to have a good eavesdropping performance. Therefore, in the first phase when S transmits information signal to R and D, E simultaneously transmits jamming noise to disrupt the suspicious communications. The received signals at R and D can be respectively expressed as $\mathbf{y}_r^{\text{JE}} = \sqrt{\frac{P_s}{d_1^\tau}} \mathbf{h}_1 x +\sqrt{\frac{1}{d_4^\tau}} \mathbf{H}_4^\dagger \mathbf{w} s+\mathbf{n}_r$, and $y_{d1}^{\text{JE}} = \sqrt{\frac{P_s}{d_2^\tau}} h_2 x+\sqrt{\frac{1}{d_6^\tau}} \mathbf{h}_6 \mathbf{w} s +n_{d1}$, where the superscript $\text{JE}$ stands for ``jamming-then-eavesdropping'', $P_s$ denotes the transmit power of S, $x$ and $s$ denote the information and jamming symbol with unit power, respectively. $\mathbf{w}$ is the $M \times 1$ transmit beamforming vector at E with $ \mathbf{w}^\dagger \mathbf{w} \leq P$, where $P$ denotes the maximum jamming power at E. In addition, $\mathbf{n}_r$ and $n_{d1}$ are the additive white Gaussian noises (AWGNs) at R and D, respectively. Without loss of generality, the elements of $\mathbf{n}_r$ and $n_{d1}$ follow zero mean Gaussian distribution with unit variance.

In the second phase, R first decodes the information from S using maximal ratio combining (MRC), and then forwards the re-encoded symbol to D using maximum ratio transmission (MRT), while E tries to overhear the signal. The received signals at D and E can be expressed as $y_{d2}^{\text{JE}} = \sqrt{\frac{P_r}{d_5^\tau}} \mathbf{h}_5 \frac{\mathbf{h}_5^\dagger}{||\mathbf{h}_5||} x+n_{d2}$, and $\mathbf{y}_e^{\text{JE}} = \sqrt{\frac{P_r}{d_4^\tau}} \mathbf{H}_4 \frac{\mathbf{h}_5^\dagger}{||\mathbf{h}_5||} x+\mathbf{n}_{e2}$, where $P_r$ denotes the transmit power of R, $n_{d2}$ and  $\mathbf{n}_{e2}$ are the AWGNs at D and E, respectively. Since D observes two copies of the signal, MRC is used to enhance the signal recovery, while E exploits the multiple antennas by using MRC for reception. Therefore, the corresponding SNRs (signal-to-interference-plus-noise ratios, SINRs) at R, D and E can be expressed as $\Gamma_{r}^{\text{JE}}=\frac{P_s d_4^\tau ||\mathbf{h}_1||^2 }{d_1^\tau \left|\frac{\mathbf{h}_1^\dagger}{||\mathbf{h}_1||} \mathbf{H}_4^\dagger \mathbf{w}\right|^2+d_1^\tau d_4^\tau N_0}$, $\Gamma_{d}^{\text{JE}} = \frac{P_r }{d_5^\tau N_0}||\mathbf{h}_5||^2+\frac{P_s d_6^\tau|{h}_2|^2}{d_2^\tau |\mathbf{h}_6 \mathbf{w}|^2+d_2^\tau d_6^\tau N_0}$, and $\Gamma_e^{\text{JE}} = \frac{P_r }{d_4^\tau N_0} ||\mathbf{H}_4 \frac{\mathbf{h}_5^\dagger}{||\mathbf{h}_5||}||^2$, respectively.

\subsection{Strategy \uppercase\expandafter{\romannumeral2}: Eavesdropping First}
In contrast to strategy \uppercase\expandafter{\romannumeral1}, it is a better choice to perform eavesdropping in the first phase if the quality of S $\rightarrow$ E link is good. In this case, S broadcasts the information signal to R and D, while E tries to eavesdrop the information. The received signals at R, D, and E can be respectively expressed as $\mathbf{y}_r^{*} = \sqrt{\frac{P_s}{d_1^\tau}} \mathbf{h}_1 x+ \mathbf{n}_r$, ${y}_{d1}^{*} = \sqrt{\frac{P_s}{d_2^\tau}} h_2 x+ {n}_{d1}$, and $\mathbf{y}_e^{*} = \sqrt{\frac{P_s}{d_3^\tau}} \mathbf{h}_3 x+ \mathbf{n}_{e1}$ with $* \in \{\text{EH},\text{EE},\text{EJ}\}$, where each superscript stands for ``eavesdropping-then-helping'', `` eavesdropping-then-eavesdropping'', `` eavesdropping-then-jamming'', respectively. Also, $\mathbf{n}_{e1}$ is the AWGN at E. Therefore, the received SNRs at R and E can be expressed as $\text{SNR}_{r}^{*}=\frac{P_s}{d_1^\tau N_0} ||\mathbf{h}_1||^2$ and $\text{SNR}_{e}^{*}=\frac{P_s}{d_3^\tau N_0} ||\mathbf{h}_3||^2$.

Now, depending on the relative channel quality of the suspicious communication link and eavesdropping link, E may take different actions in order to maximize the effective eavesdropping rate in the second phase.
\subsubsection{Helping}
If $\text{SNR}_{r}^{*} \leq \text{SNR}_{e}^{*}$, E is guaranteed to successfully decode the suspicious information. Therefore, in order to further improve the effective eavesdropping rate, E acts as a helper trying to increase the rate of the suspicious link. As such, the received signal at D can be expressed as $y_{d2}^{\text{EH}} = \sqrt{\frac{P_r}{d_5^\tau}} \mathbf{h}_5 \frac{\mathbf{h}_5^\dagger}{||\mathbf{h}_5||} x+ \sqrt{\frac{P_e}{d_6^\tau}} \mathbf{h}_6 \frac{\mathbf{h}_6^\dagger}{||\mathbf{h}_6||} x+n_{d2}$, where $P_e$ denotes the transmit power of E. For fairness of comparison between different strategies, we constrain the maximum transmit power as in strategy \uppercase\expandafter{\romannumeral1}, i.e., $0 \leq P_e \leq P$. Therefore, the corresponding SNRs can be expressed as $\Gamma_{r}^{\text{EH}} = \frac{P_s}{d_1^\tau N_0} ||\mathbf{h}_1||^2$, $\Gamma_{d}^{\text{EH}} = \frac{P_s}{d_2^\tau N_0} |h_2|^2+\frac{1}{N_0}(\sqrt{\frac{P_r}{d_5^\tau}} ||\mathbf{h}_5||+ \sqrt{\frac{P_e}{d_6^\tau}}||\mathbf{h}_6||)^2$, and $\Gamma_{e}^{\text{EH}} = \frac{P_s}{d_3^\tau N_0} ||\mathbf{h}_3||^2$, respectively.

\subsubsection{Eavesdropping}
If E is not able to decode the information in the first phase, E may choose to either continue eavesdropping the suspicious link or switch to jamming in the second phase\footnote{The mode selection criterion will be specified in the next section.}, then the received signals at D and E can be expressed as $y_{d2}^{\text{EE}} = \sqrt{\frac{P_r}{d_5^\tau}} \mathbf{h}_5 \frac{\mathbf{h}_5^\dagger}{||\mathbf{h}_5||} x +n_{d2}$, and $\mathbf{y}_e^{\text{EE}} = \sqrt{\frac{P_r}{d_4^\tau}} \mathbf{H}_4 \frac{\mathbf{h}_5^\dagger}{||\mathbf{h}_5||} x +\mathbf{n}_{e2}$. To strengthen the signal detection, E employs MRC to combine the signals from two time slots. Therefore, the corresponding SNRs can be expressed as $\Gamma_{r}^{\text{EE}}  = \frac{P_s}{d_1^\tau N_0}||\mathbf{h}_1||^2$, $\Gamma_{d}^{\text{EE}}  = \frac{P_s}{d_2^\tau N_0}|h_2|^2+\frac{P_r}{d_5^\tau N_0}||\mathbf{h}_5||^2$,
 and $\Gamma_e^{\text{EE}} = \frac{P_s}{d_3^\tau N_0}||\mathbf{h}_3||^2+\frac{P_r}{d_4^\tau N_0}||\mathbf{H}_4 \frac{\mathbf{h}_5^\dagger}{||\mathbf{h}_5||}||^2$, respectively.

\subsubsection{Jamming}
In this case, E performs jamming in an effort to degrade the rate of the suspicious link, so that the probability of the successful eavesdropping can be improved. Therefore, the received signal at D can be expressed as $y_{d2}^{\text{EJ}} = \sqrt{\frac{P_r}{d_5^\tau}} \mathbf{h}_5 \frac{\mathbf{h}_5^\dagger}{||\mathbf{h}_5||}x+\sqrt{\frac{P_e}{d_6^\tau}} \mathbf{h}_6 \frac{\mathbf{h}_6^\dagger}{||\mathbf{h}_6||}s+n_{e2}$. The corresponding SNRs (SINRs) can be expressed as $\Gamma_r^{\text{EJ}} = \frac{P_s}{d_1^\tau N_0}||\mathbf{h}_1||^2$, $\Gamma_d^{\text{EJ}} = \frac{P_s}{d_2^\tau N_0}|h_2|^2+\frac{P_r d_6^\tau ||\mathbf{h}_5||^2}{P_e d_5^\tau ||\mathbf{h}_6||^2+d_5^\tau d_6^\tau N_0}$, and $\Gamma_e^{\text{EJ}} = \frac{P_s}{d_3^\tau N_0}||\mathbf{h}_3||^2$, respectively.

\subsection{Problem Formulation}
Depending on the different strategies, the maximum achievable rate of the suspicious link and eavesdropping link can be respectively expressed as $C_{\text{SD}}=\frac{1}{2} \min \left(\log (1+\Gamma_r^\zeta),\log(1+\Gamma_d^\zeta)\right)$ and $C_{\text{SE}}= \frac{1}{2} \log(1+\Gamma_e^\zeta)$, where $\zeta \in \{\text{JE},\text{EH},\text{EE},\text{EJ}\}$. Note that the factor $\frac{1}{2}$ is due to the fact that the entire communication occupies two slots.

If $C_{\text{SE}} \geq C_{\text{SD}}$, E can reliably decode the suspicious information with arbitrarily small error probability. As such, the effective eavesdropping rate is given by $R = C_{\text{SD}}$ \cite{Y.Zeng}. On the other hand, if $C_{\text{SE}} < C_{\text{SD}}$, it is impossible for E to decode the information without any error. In this case, the effective eavesdropping rate is defined as $R=0$. Therefore, the main objective is to optimize the transmit beamforming vector $\mathbf{w}$ for strategy \uppercase\expandafter{\romannumeral1} or the transmit power $P_e$ for strategy \uppercase\expandafter{\romannumeral2} at E, so that the effective eavesdropping rate is maximized. The corresponding optimization problem can be formulated as
\begin{align} \label{10}
\text{(P1)} \quad  \quad  {\mathop{\max }} \quad & C_{\text{SD}}  \nonumber \\
\text{s.t.} \quad & C_{\text{SE}}  \geq  C_{\text{SD}}  \nonumber\\
& \mathbf{w}^\dagger \mathbf{w} \leq P \quad \text{or} \quad  0 \leq P_e \leq P.
\end{align}

\section{Optimal Beamformer and Transmit Power Design}
In this section, we study the optimal beamformer and transmit power design of optimization problem (P1).
\subsection{Strategy \uppercase\expandafter{\romannumeral1}: Jamming first}
Since logarithm is a monotonically increasing function, problem (P1) can be reformulated as
\begin{align}
\text{(P2)} \quad \underset{ \mathbf{w}}{\mathop{\max }} \quad & f(\mathbf{w})   \nonumber \\
\text{s.t.} \quad &\Theta_6 \geq f(\mathbf{w}) \quad  \text{and} \quad \mathbf{w}^\dagger \mathbf{w} \leq P,
\end{align}
where $f(\mathbf{w})=\min ( \Theta_1+\frac{\Theta_2}{|\mathbf{h}_6 \mathbf{w}|^2+\Theta_3},\frac{\Theta_4}{ |\frac{\mathbf{h}_1^\dagger}{||\mathbf{h}_1||} \mathbf{H}_4^\dagger \mathbf{w}|^2+\Theta_5} )$, $\Theta_1 = \frac{P_r }{d_5^\tau N_0} ||\mathbf{h}_5||^2 $, $\Theta_2 = \frac{P_s d_6^\tau}{d_2^\tau} |h_2|^2$, $\Theta_3 = d_6^\tau N_0$, $\Theta_4 =  \frac{P_s d_4^\tau}{d_1^\tau} ||\mathbf{h}_1||^2 $, $\Theta_5 = d_4^\tau N_0$, and $\Theta_6 = \frac{P_r}{d_4^\tau N_0} ||\mathbf{H}_4 \frac{\mathbf{h}_5^\dagger}{||\mathbf{h}_5||}||^2$. It is easy to observe that the maximum value of $f\left(\mathbf{w}\right)$ equals to $f_{\max} \left(\mathbf{w}\right) = \min (\Theta_1+\frac{\Theta_2}{\Theta_3},\frac{\Theta_4}{\Theta_5})$, which can be achieved when $\mathbf{w}=\mathbf{0}$. Then we have the following important observation:
\begin{lemma}
The optimal transmit beamformer that minimizes $f\left(\mathbf{w}\right)$ can be expressed as
\begin{align}
\mathbf{w}_{\text{opt}} = \sqrt{x} \frac{\Pi_{\mathbf{h}_6^\dagger} \mathbf{H}_4 \mathbf{h}_1}{||\Pi_{\mathbf{h}_6^\dagger} \mathbf{H}_4 \mathbf{h}_1||} + \sqrt{P-x} \frac{\Pi_{\mathbf{h}_6^\dagger}^{\perp} \mathbf{H}_4 \mathbf{h}_1}{||\Pi_{\mathbf{h}_6^\dagger}^{\perp} \mathbf{H}_4 \mathbf{h}_1||},
\end{align}
with $0\leq x \leq P$.
\proof The proof of Lemma 1 can be referred to \cite{J.Y.Ryu}.
\end{lemma}
Substituting $\mathbf{w}_{\text{opt}}$ into $f\left(\mathbf{w}\right)$, and define $g\left(x\right) = \min (\Theta_1+ \\  \frac{\Theta_2}{x ||\mathbf{h}_6||^2+\Theta_3}, \frac{\Theta_4}{(\sqrt{x}\frac{||\Pi_{\mathbf{h}_6^\dagger} \mathbf{H}_4 \mathbf{h}_1||}{||\mathbf{h}_1||}+\sqrt{P-x}\frac{||\Pi_{\mathbf{h}_6^\dagger}^{\perp} \mathbf{H}_4 \mathbf{h}_1||}{||\mathbf{h}_1||})^2+\Theta_5})$. It is easy to show that the minimum value of $g\left(x\right)$ can be achieved when $x = P$ or $x = \frac{||\Pi_{\mathbf{h}_6^\dagger} \mathbf{H}_4 \mathbf{h}_1||^2}{||\mathbf{H}_4 \mathbf{h}_1||^2}P$, and we have $f_{\min} \left(\mathbf{w}\right)= \min (\Theta_1+\frac{\Theta_2}{P||\mathbf{h}_6||^2+\Theta_3},\frac{\Theta_4}{P \frac{||\mathbf{H}_4 \mathbf{h}_1||^2}{||\mathbf{h}_1||^2}+\Theta_5})$.

We now consider three separate cases:\\
1) If $\Theta_6 > f_{\max} \left(\mathbf{w}\right)$, i.e., the eavesdropping channel is sufficiently good that E is able to decode the information successfully without any help. As such, it is better for E to eavesdrop rather than to jam, i.e., $\mathbf{w}=\mathbf{0}$. Therefore, the corresponding eavesdropping rate is $\frac{1}{2} \log \left(1+f_{\max} \left(\mathbf{w}\right)\right)$. \\
2) If $\Theta_6 < f_{\min} \left(\mathbf{w}\right)$, i.e., the eavesdropping channel is too weak that even jamming with full-power is insufficient, thus there is no need to waste jamming power. Therefore, we set $\mathbf{w}=\mathbf{0}$, the resulting eavesdropping rate is 0. \\
3) Otherwise, since $f \left(\mathbf{w}\right)$ is a continuous function of $\mathbf{w}$, there exists $\mathbf{w}$ satisfying $f \left(\mathbf{w}\right)=\Theta_6$. As such, the corresponding eavesdropping rate is $\frac{1}{2} \log \left(1+\Theta_6\right)$, where the optimal $\mathbf{w}$ can be obtained with the help of semidefinite programming (SDP) technique \cite{S.Boyd}. Specifically, let $\mathbf{W} = \mathbf{w} \mathbf{w}^\dagger$ and ignore the rank-one constraint, $\mathbf{W}$ can be found by
\begin{align}
\text{(P3)}   \quad \underset{ \mathbf{W}}{\mathop{\min }} \quad &   0   \\
\text{s.t.} \quad   & \text{tr} (\mathbf{W} \mathbf{h}_{6}^\dagger \mathbf{h}_{6})=\frac{\Theta_2}{\Theta_6-\Theta_1}-\Theta_3,  \nonumber \\
& \text{tr} (\mathbf{W} \mathbf{H}_{4} \mathbf{h}_{1} \mathbf{h}_1^\dagger \mathbf{H}_4^\dagger ) \leq ||\mathbf{h}_1||^2 (\frac{\Theta_4}{\Theta_6}-\Theta_5), \nonumber \\
& \text{tr} (\mathbf{W}) \leq P,  \qquad \mathbf{W} \succeq \mathbf{0}, \nonumber
\end{align}
or
\begin{align}
\text{(P4)}   \quad \underset{ \mathbf{W}}{\mathop{\min }} \quad &   0   \\
\text{s.t.} \quad
& \text{tr} (\mathbf{W} \mathbf{H}_{4} \mathbf{h}_{1} \mathbf{h}_1^\dagger \mathbf{H}_4^\dagger ) = ||\mathbf{h}_1||^2 (\frac{\Theta_4}{\Theta_6}-\Theta_5), \nonumber \\
& \text{tr} (\mathbf{W} \mathbf{h}_{6}^\dagger \mathbf{h}_{6}) \leq \frac{\Theta_2}{\Theta_6-\Theta_1}-\Theta_3,  \nonumber \\
& \text{tr} (\mathbf{W}) \leq P,  \qquad \mathbf{W} \succeq \mathbf{0}. \nonumber
\end{align}
Note that at least one of the optimization problems is feasible and the convex SDP problem can be efficiently solved using the CVX tools \cite{S.Boyd}. Due to the fact that the optimal solution may have a rank higher than one, we need to use some approximation approaches such as randomization to find the approximate beamforming vectors \cite{N.D.Sidiropoulos}.

\subsection{Strategy \uppercase\expandafter{\romannumeral2}: Eavesdropping First}
In this subsection, we provide the optimal transmit power solution of problem (P1).

If $\text{SNR}_r^* \leq \text{SNR}_e^*$, E can always successfully decode the suspicious message. Hence, the effective eavesdropping rate is determined by the maximum achievable rate of the suspicious link. Since the achievable rate of the dual-hop suspicious link is limited by the rate of the weakest hop, E may act as a helper to improve the SNR of the second hop if it is inferior to the SNR of the first hop. Depending on the effective SNRs of the two hop channels, we consider three different cases:\\
1) If $\frac{P_s}{d_1^\tau N_0}||\mathbf{h}_1||^2 \leq \frac{P_s}{d_2^\tau N_0}|h_2|^2+\frac{P_r}{d_5^\tau N_0}||\mathbf{h}_5||^2$, the effective SNR of the second hop without help from E is larger than that of the first hop, then E can remain silent, i.e., $P_e = 0$, and the corresponding eavesdropping rate is $\frac{1}{2} \log (1+\frac{P_s}{d_1^\tau N_0}||\mathbf{h}_1||^2)$. \\
2) If $\frac{P_s}{d_1^\tau N_0}||\mathbf{h}_1||^2 \geq \frac{1}{N_0}(\sqrt{\frac{P_r}{d_5^\tau}} ||\mathbf{h}_5||+ \sqrt{\frac{P}{d_6^\tau}}||\mathbf{h}_6||)^2+\frac{P_s}{d_2^\tau N_0}|h_2|^2$, the effective SNR of the second hop is too weak that E should help with maximum transmit power $P$, and the eavesdropping rate is $\frac{1}{2} \log(1+\frac{P_s}{d_2^\tau N_0}|h_2|^2+\frac{1}{N_0}(\sqrt{\frac{P_r}{d_5^\tau}} ||\mathbf{h}_5||+ \sqrt{\frac{P}{d_6^\tau}}||\mathbf{h}_6||)^2)$. \\
3) Otherwise, E can adjust its transmit power to maintain $\Gamma_{r}^{\text{EH}}=\Gamma_{d}^{\text{EH}}$, i.e., $P_e = \frac{d_6^\tau}{||\mathbf{h}_6||^2} (\sqrt{\frac{P_s}{d_1^\tau}||\mathbf{h}_1||^2-\frac{P_s}{d_2^\tau}|h_2|^2}-\sqrt{\frac{P_r}{d_5^\tau}} ||\mathbf{h}_5||)^2$, and the corresponding eavesdropping rate is  $\frac{1}{2} \log (1+\frac{P_s}{d_1^\tau N_0}||\mathbf{h}_1||^2)$.

If $\text{SNR}_r^* > \text{SNR}_e^*$, E is not guaranteed to decode the suspicious information successfully in the first phase, thus E can act as an eavesdropper or a jammer in the second phase. Three cases are studied: \\
1) If $\frac{P_s}{d_3^\tau N_0} ||\mathbf{h}_3||^2 \geq \frac{P_s}{d_2^\tau N_0} |h_2|^2+\frac{P_r}{d_5^\tau N_0} ||\mathbf{h}_5||^2$, the effective SNR of the eavesdropping channel is larger than that of the second hop channel, then E can always successfully decode the suspicious information. Therefore, E can remain silent in the second phase, and the corresponding eavesdropping rate is $\frac{1}{2} \log (1+\frac{P_s}{d_2^\tau N_0} |h_2|^2+\frac{P_r}{d_5^\tau N_0} ||\mathbf{h}_5||^2)$. \\
2) If $ \frac{P_s}{d_3^\tau N_0} ||\mathbf{h}_3||^2 < \frac{P_s}{d_2^\tau N_0}|h_2|^2+\frac{P_r d_6^\tau ||\mathbf{h}_5||^2}{P d_5^\tau ||\mathbf{h}_6||^2+d_5^\tau d_6^\tau N_0}$, the effective SNR of the eavesdropping channel is too weak such that even full-power jamming does not work, E has no choice but to eavesdrop again to enhance the eavesdropping performance. Thus, the eavesdropping rate is
$\frac{1}{2} \log (1+\min (\frac{P_s}{d_1^\tau N_0}||\mathbf{h}_1||^2,\frac{P_s}{d_2^\tau N_0} |h_2|^2+\frac{P_r}{d_5^\tau N_0}||\mathbf{h}_5||^2))$ when $\Gamma_e^{\text{EE}} \geq \min \left(\Gamma_r^{\text{EE}},\Gamma_d^{\text{EE}}\right)$, otherwise the eavesdropping rate is 0. \\
3) Otherwise, E can select to perform eavesdropping or jamming to ensure the success of decoding. Since $\frac{P_s}{d_1^\tau N_0}||\mathbf{h}_1||^2>\frac{P_s}{d_3^\tau N_0} ||\mathbf{h}_3||^2$ and $\frac{P_s}{d_2^\tau N_0} |h_2|^2+\frac{P_r}{d_5^\tau N_0}||\mathbf{h}_5||^2>\frac{P_s}{d_3^\tau N_0} ||\mathbf{h}_3||^2$, which indicates that eavesdropping provides a higher rate, we propose to use E as an eavesdropper if $\Gamma_e^{\text{EE}} \geq \min \left(\Gamma_r^{\text{EE}},\Gamma_d^{\text{EE}}\right)$, and the corresponding eavesdropping rate is $\frac{1}{2} \log (1+\min (\frac{P_s}{d_1^\tau N_0}||\mathbf{h}_1||^2,\frac{P_s}{d_2^\tau N_0} |h_2|^2+\frac{P_r}{d_5^\tau N_0}||\mathbf{h}_5||^2))$. If $\Gamma_e^{\text{EE}} < \min \left(\Gamma_r^{\text{EE}},\Gamma_d^{\text{EE}}\right)$, E can adjust its jamming power to maintain $\Gamma_{e}^{\text{EJ}}=\Gamma_{d}^{\text{EJ}}$, i.e., $P_e = \frac{d_6^\tau N_0}{d_5^\tau ||\mathbf{h}_6||^2} (\frac{P_r d_2^\tau d_3^\tau ||\mathbf{h}_5||^2}{P_s d_2^\tau ||\mathbf{h}_3||^2-P_s d_3^\tau |h_2|^2}-d_5^\tau)$, and the resulting eavesdropping rate is $\frac{1}{2} \log (1+\frac{P_s}{d_3^\tau N_0} ||\mathbf{h}_3||^2)$.

\section{Numerical Results}
In this section, we present numerical results to evaluate the performance of the proposed proactive strategies. Unless otherwise specify, we set the carrier frequency as 5 GHz, the bandwidth as 20 MHz, the noise power density as -174 dBm/Hz, and the transmit power of S and R as $P_s=P_r=40$ dBm. The path-loss exponent is $\tau=3$, and the numbers of the antennas at R and E are $N=M=4$. The four nodes are located in a 2-D topology, where S, R, D, and E are located at $(0,0)$ km, $(2,0)$ km, $(4,0)$ km, and $(2,3)$ km, respectively.

\begin{figure}[htbp]
\centering
\includegraphics[width=2.1in]{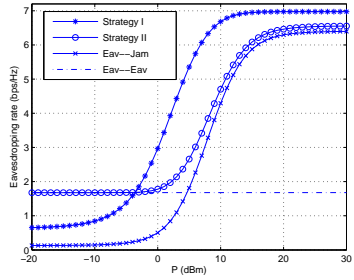}
\caption{Eavesdropping rate comparison of the two strategies and other benchmark schemes.}
\label{fig:fig2}
\end{figure}

Fig. \ref{fig:fig2} depicts the achievable eavesdropping rate of different strategies. For comparison, the performance of two heuristic benchmark schemes are also plotted, namely, eavesdropping-then-jamming and eavesdropping-then-eavesdropping. As expected, strategy \uppercase\expandafter{\romannumeral2} always outperforms the two reference schemes, since it adaptively adjusts the action of the legitimate monitor to enhance the eavesdropping rate of the system. However, the performance difference of strategy \uppercase\expandafter{\romannumeral1} and \uppercase\expandafter{\romannumeral2} depends heavily on the network topology and operating parameters. In the current setup, strategy \uppercase\expandafter{\romannumeral1} is inferior in the low jamming power regime, while becomes superior when there is sufficient available jamming power.

\begin{figure}[htbp]
\centering
\includegraphics[width=2.1in]{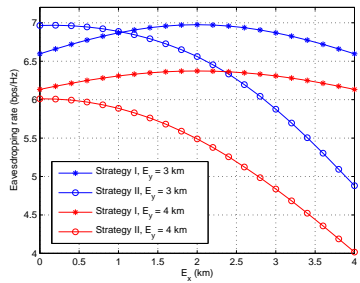}
\caption{Eavesdropping rate versus different positions of the legitimate monitor with $P=40$ dBm.}
\label{fig:fig3}
\end{figure}

In Fig. \ref{fig:fig3}, we plot the eavesdropping rate as a function of the legitimate monitor's location for the two strategies where the y-coordinate ($\text{E}_{\text{y}} $) of E is fixed as 3 km or 4 km, while the x-coordinate ($\text{E}_{\text{x}}$) of E varies within the range of [0,4] km. We observe that when E moves away from S, the eavesdropping rate of strategy \uppercase\expandafter{\romannumeral2} monotonically decreases. This is intuitive since the eavesdropping performance degrades when the distance between E and S increases. In contrast, there is an optimum position for E when adopting strategy \uppercase\expandafter{\romannumeral1}, i.e., the point that is most close to R. This reason is that when E is near R, it can perform jamming and eavesdropping efficiently. Therefore, as a simple criterion, E employs strategy \uppercase\expandafter{\romannumeral2} when it is close to S, and employs strategy \uppercase\expandafter{\romannumeral1} when it is close to R.

\section{Conclusion}
This paper considered the issue of legitimate surveillance in a dual-hop DF relaying system. Specifically, two strategies aiming at maximizing the effective eavesdropping rate have been proposed, and the corresponding optimal beamformer and power allocation scheme have been obtained. It was shown that the proposed strategies significantly outperform other benchmark schemes. Moreover, the position of the legitimate monitor can be used as a simple criterion to determine the appropriate strategy.

\newpage

\nocite{*}
\bibliographystyle{IEEE}
\begin{footnotesize}

\end{footnotesize}

\end{document}